\documentclass[11pt]{article}
\usepackage{latexsym,amsthm,amssymb,amsfonts,amsbsy,graphicx,lscape,array,citesort,multibox}

\def\d{\delta}

\def\l{\lambda}

\def\m{\mu}
\def\n{\nu}
\def\r{\rho}
\def\o{\omega}
\def\O{\Omega}
\def\s{\sigma}

\def\pa{\partial}

\def\be{\begin{equation}}
\def\ee{\end{equation}}
\def\beq{\begin{eqnarray}}
\def\eeq{\end{eqnarray}}

\def\co{{\cal O}}

\newcommand{\bqn}{\begin{eqnarray}}\newcommand{\eqn}{\end{eqnarray}}
\begin{document}
\begin{flushright}
IHES/P/05/54 \\
ULB-TH/06-04
\end{flushright}

\vspace{.3cm}

\begin{centering}

{\Large {\bf On Killing tensors and cubic vertices in higher-spin
gauge theories}}

\vspace{.5cm}

\begin{center}
{Xavier Bekaert$^{a,}$\footnote{E-mail address:
\tt{bekaert@ihes.fr}}, Nicolas Boulanger$^{b,}$\footnote{Charg\'e
de Recherches FNRS (Belgium); \tt{nicolas.boulanger@umh.ac.be}},
Sandrine Cnockaert$^{c,}$\footnote{Aspirant du FNRS (Belgium);
\tt{sandrine.cnockaert@ulb.ac.be}} and
Serge Leclercq$^{b,}$\footnote{E-mail address: {\tt serge.leclercq@umh.ac.be}} }
\end{center}

{\small{
\begin{center}
$^a$
Institut des Hautes \'Etudes Scientifiques\\
Le Bois-Marie, 35 route de Chartres, 91440 Bures-sur-Yvette
(France)\\
\vskip1mm $^b$
Universit\'e de Mons-Hainaut, M\'ecanique et Gravitation \\
6 avenue du Champ de Mars, 7000 Mons (Belgium)\\
\vskip1mm $^c$ Physique Th\'eorique et Math\'ematique,
Universit\'e Libre
de Bruxelles\\
and International Solvay Institutes,\\
U.L.B. Campus Plaine, C.P. 231, B-1050, Bruxelles (Belgium)\\
\end{center}}}
\end{centering}
\vspace*{.2cm}
{\textit{Based on a talk given by X.B. at the RTN Workshop ``Constituents,
Fundamental Forces and Symmetries of the Universe'' (Corfu, 20-26th
September 2005). Contribution to the Proceedings.}}
\vspace*{.2cm}
\begin{abstract}
The problem of determining all consistent non-Abelian local
interactions is reviewed in flat space-time. The antifield-BRST
formulation of the free theory is an efficient tool to address
this problem. Firstly, it allows to compute all on-shell local
Killing tensor fields, which are important because of their deep
relationship with higher-spin algebras. Secondly, under the sole
assumptions of locality and Poincar\'e invariance, all non-trivial
consistent deformations of a sum of spin-three quadratic actions
deforming the Abelian gauge algebra were determined. They are
compared with lower-spin cases.
\end{abstract}

\section{Introduction}

It has been known since the works of Wigner that first-quantized
relativistic particles are in one-to-one correspondence with
inequivalent unitary irreducible representations (UIRs) of the
Poincar\'e group $ISO(n-1,1)$. Therefore it is a natural programme
to explore the exhaustive list of such possibilities and to
determine if some exotic cases could be of physical relevance. At
sufficiently high energy, any given massive field effectively
behaves as a massless field. The little group\footnote{In order
for the little group to make sense, we will consider here
space-time dimension $n\geqslant 3$.} $ISO(n-2)$ of a light-like
momentum governs the classification of the massless
representations which fall into two distinct categories: the
particularly interesting ``helicity" (i.e. finite-dimensional)
representations and the generic ``continuous spin" (i.e.
infinite-dimensional) representations. The helicity fields are
either the celebrated completely symmetric gauge fields (see e.g.
\cite{Intros} for introductions) or the mixed symmetry gauge
fields (see for instance \cite{Dubna} for a review). Usually the
``spin" $s$ of a single-valued helicity representation refers to
the number of columns of the Young diagram labeling the
representation of the maximal compact subgroup $SO(n-2)$. For
completely symmetric tensor gauge fields, the spin $s$ is equal to
the rank. ``Lower spin" stands for spin $s\leqslant 2\,$, while
``higher spin" refers to spin $s>2\,$. Covariant field equations
were recently obtained for each continuous spin representation
from a subtle infinite spin limit of helicity field equations
\cite{Bekaert:2005in} so that ``continuous spin" may somehow be
thought of as the case $s=\infty\,$.

Whereas gauge theories describing free massless fields are by now
well established, it still remains unclear whether non-trivial
consistent self-couplings and/or cross-couplings among those
fields may exist in general, such that the deformed gauge algebra
is non-Abelian. The goal of the present paper is to review a
mathematically precise statement of this problem and to focus on
the rather general results obtained thanks to a very useful tool:
the Becchi-Rouet-Stora-Tyutin (BRST) reformulation of the Noether
procedure for the determination of the cubic vertices and
corresponding deformations of the gauge transformations.

\section{The interaction problem}

We review the deformation setting for the problem of constructing
consistent local vertices for a free field theory possessing some
gauge symmetries \cite{Noether} and particularize to the case of
non-Abelian deformations. The cohomological reformulation of the
deformation problem in the antifield-BRST formalism
\cite{Barnich:1993vg} is briefly sketched. Note that other
formulations exist (see e.g. \cite{AncoDefo} and refs therein).

\subsection{General hypotheses}

The starting point is some given action $S_0$ that is said to be
``undeformed". We assume, as in the Noether deformation procedure,
that the deformed action can be expressed as a formal power series
in some coupling constants denoted collectively by $g\,$, the
zeroth-order term in the expansion describing the undeformed
theory $S_0\,$:$$S=S_0+g\,S_1+g^2S_2+\co(g^3)\,.$$ The procedure
is then p\underline{erturbative}: one tries to construct the
deformations order by order in the deformation parameters $g\,$.
Some physical requirements naturally come out:
\begin{itemize}
  \item \underline{Poincar\'e }sy\underline{mmet}ry: We ask that
the Lagrangian be {\it manifestly} invariant under the
{\em{Poincar\'e}} group. Therefore, the free theory is formulated
in the flat space-time ${\mathbb R}^{n-1,1}$, the Lagrangian
should not depend explicitly on the space-time cartesian
coordinates $\{x^\mu\}$ and all space-time indices must be raised
and lowered by using either the Minkowski metric or the
Levi-Civita tensor.
  \item \underline{non-triviali}ty: We reject {\em trivial} deformations
arising from field-redefinitions that reduce to the identity at
order zero:
$$
\varphi\,\longrightarrow\, \varphi'=\varphi+g\, \phi (\varphi, \pa
\varphi, \cdots)+\co(g^2)\,.$$
  \item \underline{Consisten}cy: A deformation of a
theory is called {\em{consistent}} if the deformed theory
possesses the same number of possibly deformed (perturbatively as
well) independent gauge symmetries, reducibility identities, {\it
etc.}, as the system we started with. In other words, the number
of physical degrees of freedom is unchanged.
  \item \underline{Locali}ty: The deformed action $S[\varphi]$ must be a {\em local}
  functional. The deformations of the gauge transformations, {\it etc.},
  must be local functions, as well as the allowed field redefinitions.
\end{itemize}
We remind the reader that a local function of some set of fields
$\varphi$ is a smooth function of the fields $\varphi$ and their
derivatives $\partial\varphi$, $\partial^2\varphi$, ... up to some
{\it finite} order, say $k$, in the number of derivatives. Such a
set of variables $\varphi$, $\partial\varphi$, ...,
$\partial^k\varphi$ will be collectively denoted by $[\varphi]$.
Therefore, a local function of $\varphi$ is denoted by
$f([\varphi])$. A local $p$-form $(0\leqslant p \leqslant n)$ is a
differential $p$-form the components of which are local functions:
\begin{eqnarray}
        \omega= \frac{1}{p!}\,\omega_{\m_1\ldots\m_p}(x, [\varphi])\, dx^{\m_1}
\wedge \cdots \wedge dx^{\m_{p}}\,.\nonumber
\end{eqnarray}
A local functional $\O[\varphi]$ of the fields $\varphi$ is the
integral $\O=\int \o$ of a local $n$-form $\o=\o(x,
[\varphi])$.\vspace{1mm}

The general problem of consistently deforming a given action $S_0$
is of course highly non-trivial. It is very fruitful to
investigate it order by order in $g$ and to exploit the undeformed
gauge symmetries that strongly restrict the consistent
possibilities. Indeed, one can easily show that non-trivial first
order consistent local deformations $S_1$ are, on-shell,
non-vanishing gauge invariant local functionals (with respect to
the undeformed equations of motion and gauge symmetries). To
reformulate the problem in the antifield-BRST setting it is enough
to observe \cite{Barnich:1993vg} that the latter functionals are
in one-to-one correspondence with elements of the local BRST
cohomology group $H^{n,0}(s_0|\,d)$ in top form degree and in
vanishing ghost number, where $s_0$ is the BRST differential
corresponding to the undeformed action $S_0$. Moreover, the
obstructions to the existence of a second-order deformation $S_2$
corresponding to $S_1$ are encoded in the local BRST cohomology
group $H^{n,1}(s_0|\,d)$ in ghost number one
\cite{Barnich:1993vg}.

\subsection{Non-Abelian deformations}

The conventional local free theories corresponding to UIRs of the
helicity group $SO(n-2)$ that are completely symmetric tensors
have been constructed by Fronsdal a while ago
\cite{Fronsdal:1978rb} for arbitrary rank $s$. To have Lorentz
invariance manifest, the theory is expressed in terms of
completely symmetric tensor gauge fields
$\varphi_{\m_1\ldots\,\,\m_s}=\varphi_{(\m_1\ldots\,\,\m_s)}$ of
rank $s>0$, the gauge transformation of which reads
$$
\delta_\varepsilon\varphi_{\m_1\ldots\,\,\m_s}=s\,\partial_{(\m_1}\varepsilon_{\m_2\ldots\m_s)}\,,
$$
where the curved (square) bracket denotes complete
(anti)symmetrization with strength one\footnote{For example,
$\Phi_{(\m\n)}=\frac{1}{2}(\Phi_{\m\n}+\Phi_{\n\m})$ and
$\Phi_{[\m\n]}=\frac{1}{2}(\Phi_{\m\n}-\Phi_{\n\m})$.} and the
Greek indices run over $n$ values ($n\geqslant 3$). The gauge
parameter $\varepsilon$ is a completely symmetric tensor field of
rank $s-1$. For spin $s=1$ the gauge field $\varphi_\m$ represents
the photon with $U(1)$ gauge symmetry while for spin $s=2$ the
gauge field $\varphi_{\mu\nu}$ represents the graviton with
linearized diffeomorphism invariance. The gauge field theories
corresponding to tensorial helicity representations labeled by
one-column Young diagrams are the usual $p$-form ({\it i.e.}
completely antisymmetric tensor) gauge theories. Analogous gauge
field theories corresponding to arbitrary spin-two ({\it i.e.}
two-column Young diagrams) helicity representations \cite{C+AKO}
have been studied recently \cite{deMedeiros:2002ge,BBC} by using
multiform and hyperform calculus \cite{Olver}.

Let us denote by $S_0[\varphi_Y]$ the Poincar\'e-invariant, local,
second-order, quadratic, gauge-invariant ghost-free actions
mentioned above where the Young diagram $Y$ labels the
corresponding representations. Now, the interaction problem
reviewed here can be formulated in a mathematically precise way as
follows:

\vspace{1mm}\noindent{\bfseries{Non-Abelian interaction problem:}}
\textit{List all Poincar\'e-invariant, non-trivial consistent
local deformations
$$S[\varphi]=S_0[\varphi]+g\,S_1[\varphi]+g^2\,S_2[\varphi]+\co(g^3)$$ of a finite positive sum
$$S_0[\varphi]=\sum_{Y,\,a} S_0[\varphi^a_Y]$$ of a collection $\varphi\equiv \varphi^a_Y$ (labeled by some index $a$ for each given Young diagram $Y$) of free gauge
field theories such that the deformed local gauge transformations
$$\d\varphi^a =\d_0\varphi^a +g\,F^a([\varphi^b],[\varepsilon^c]) +\co(g^2)$$ are \underline{non-Abelian at} f\underline{irst order} in the coupling constants
$g$.}

\vspace{1mm}Of course, this problem is too complicated to be
addressed in full generality with the techniques known at the
moment. The restriction of the interaction problem to symmetric
tensor gauge fields of rank two, is sometimes referred to as the
``Gupta programme". The generalization of the latter to a
collection of symmetric tensor gauge fields with arbitrary values
of the rank $s$ was proposed in \cite{Fronsdal:1978rb} and is
thereby frequently called the ``Fronsdal programme" or
``higher-spin interaction problem".

\subsection{Fronsdal's programme}

This old programme is still far away from completion even though
encouraging progresses have been obtained over the years. On the
one hand, the problem of consistent interactions among only
higher-spin gauge fields (hence without gravity) in Minkowski
space-time ${\mathbb R}^{n-1,1}$ was addressed in
\cite{Bengtsson:1983pd,Berends:1984wp,Bengtsson:1983bp,Bekaert:2005jf,BCL}
(and refs therein) where some positive results have been obtained
at first order in the perturbation. In the light-cone gauge,
three-point couplings between completely symmetric gauge fields
with arbitrary spins $s>2$ were constructed in
\cite{Bengtsson:1983pd}. For the pure spin-$3$ case, a cubic
vertex was obtained in a covariant form by Berends, Burgers and
van Dam (BBvD) \cite{Berends:1984wp} while new explicit vertices
were obtained very recently in \cite{Bekaert:2005jf,BCL}. The BBvD
interaction, however, leads to inconsistencies when pushed at the
next orders in powers of $g$, as was demonstrated in
\cite{Bengtsson:1983bp,Berends:1984wp,Bekaert:2005jf}. On the
other hand, the first explicit attempts to introduce minimal
coupling between higher-spin gauge fields and gravity encountered
severe problems \cite{diff}. Very early, the idea was proposed
that a consistent higher-spin gauge theory could exist, provided
all spins are taken into account \cite{Fronsdal:1978rb}. In order
to overcome the gravitational coupling problem, it was also
suggested to perturb around a maximally-symmetric curved
background, like for example $AdS_n$, in which directions
interesting results have indeed been obtained, such as cubic
vertices consistent at first order \cite{Fradkin:1987ks} and
equations of motion formally consistent at all orders
\cite{Vasiliev:2004qz} (see also \cite{Sagnotti:2005} and refs
therein).

If there is a lesson to learn from decades of efforts on the
higher-spin interaction problem, it certainly is the unusual
character of the possible interactions. For instance, the cubic
vertices contain more than two derivatives.\footnote{The full
non-linear higher-spin theory exposed in \cite{Vasiliev:2004qz}
is even expected to be non-local.}
In order to remove any prejudice on the form of the interactions,
it is natural to attack the Fronsdal programme on exhaustive and
purely algebraic grounds such as the antifield-BRST deformation
procedure.

\section{Killing tensor fields}

A problem of physical interest for a better understanding of the
higher-spin symmetries is the determination of all Killing tensor
fields on Minkowski space-time, that is, the symmetric tensor
fields satisfying the following Killing-like equation
$\partial_{(\m_1}\varepsilon_{\m_2\ldots\m_{s})}(x)=0$,
so that the corresponding Abelian gauge transformations
vanish: $\d_\varepsilon\varphi=0\,$.
The most general smooth solution of this equation is
\begin{eqnarray}
\varepsilon_{\m_1\ldots\,\m_{s-1}}(x)=\sum\limits_{t=0}^{s-1}
\l_{\m_1\ldots\,\m_{s-1}\,,\,\,\n_1\ldots\,\n_t}\,x^{\n_1}\ldots
\,x^{\n_t}\,,\quad\l_{(\m_1\ldots\,\m_{s-1}\,,\,\,\n_1)\n_2\ldots\,\n_t}=0\,
\end{eqnarray}
where each coefficient of the term of given homogeneity degree in
the coordinates $\{x^\m\}$ is a constant tensor
$\l_{\m_1\ldots\,\m_{s-1}\,,\,\,\n_1\ldots\,\n_t}$, the symmetries
of which are labeled by a two-row Young diagram.

Another motivation is that non-trivial on-shell local Killing
tensor fields are in one-to-one correspondence with cocycles of
the local Koszul-Tate cohomology group $H^n_2(\delta_0|d)$ in top
form degree and antifield number two, the knowledge of which is an
important ingredient in the computation of the local BRST
cohomology group $H^{n,0}(s_0|d)$.

\vspace{1mm}\noindent{\bfseries{Constant-curvature space-time
Killing tensors}} \cite{Bekaert:2005ka,Barnich:2005bn}:
\textit{All on-shell Killing tensor fields
$\varepsilon_{\m_1\ldots\,\m_{s-1}}(x,[\varphi])$ of the
completely symmetric tensor gauge field theory on
constant-curvature space-times can be represented by off-shell
Killing tensor fields that are independent of the gauge field
$\varphi$ and that are solutions of the Killing-like equation
$\nabla_{(\m_1}\varepsilon_{\m_2\ldots\m_{s})}(x)=0$.}

Generally speaking, the global symmetries of a solution of some
field equation correspond to the space of gauge parameters leaving
the gauge fields invariant under gauge transformations evaluated
at the solution. Furthermore, for the flat vacuum solution they
are expected to correspond to the full rigid symmetry algebra of
the theory. More specifically, the Minkowski Killing tensors of
the infinite tower of higher-spin fields should be related to a
higher-spin algebra in flat space-time.

The higher-spin gauge symmetry algebras might eventually find
their origin in the general procedure of ``gauging" some global
higher-symmetry algebras of free theories, as was argued in
\cite{Bekaert:2005ka,Modave} and as we briefly sketch here. All
linear relativistic wave equations $\textsc{K}|\phi\rangle=0$
(corresponding to some finite-dimensional UIR of the little group)
can be derived from an action taking the form of an inner product
$\int \langle\phi |\textsc{K}|\phi\rangle$. Let $\{\textsc{T}_i\}$
be Hermitian operators spanning some symmetry Lie algebra. This
means that they commute with the kinetic operator so that
$\{i\textsc{T}_i\}$ generate, via exponentation, unitary operators
preserving the quadratic action. But exactly the same is true for
any Weyl-ordered polynomial $P(\textsc{T}_i)$ of such symmetry
generators so that the symmetry algebra may become
infinite-dimensional. If the symmetry algebra is a
finite-dimensional internal algebra then the latter procedure does
not produce anything interesting in general. The case of interest
is when one deals with a space-time symmetry algebra generated by
vector fields. In such a case, the polynomials in the basis
elements are differential operators and their exponentiation leads
to non-local unitary operators \cite{Modave}.

\vspace{1mm}\noindent{\bfseries{Minkowski higher-spin algebra}}
\cite{Bekaert:2005ka}: \textit{The algebra of Weyl-ordered
polynomials in the Killing vector fields $\partial_\mu$ and
$x_{[\m}\partial_{\n]}$ of Minkowski space-time is isomorphic to
the algebra of differential operators given by
$\varepsilon_{\m_1\ldots\,\m_{s-1}}(x)\,\partial^{\m_1}\ldots\partial^{\m_{s-1}}$
defined by the infinite tower of Minkowski Killing tensor fields
($0<s<\infty$).}\vspace{1mm}

{}From its definition, it follows directly that this Minkowski
higher-spin algebra can also be obtained via an
In\"{o}n\"{u}-Wigner contraction of the $(A)dS_n$ higher-spin
algebras of Vasiliev \cite{Vasiliev:2004qz} in the flat limit
$\Lambda\rightarrow 0$. To end this section, we underline that we
have not discussed at all here the subtle issue of trace
conditions and their relation with the factorization of the
higher-spin algebras which has been debated recently
\cite{Sagnotti:2005} (in the specific context of Minkowski
Killing tensors, it was also discussed in \cite{Bekaert:2005ka}).

\section{Non-Abelian gauge transformations}

The results on one one-column \cite{Barnich:1993pa} and on
two-column \cite{Boulanger:2000rq,BBC}
Young-diagram gauge fields
together with the spin-three case \cite{Bekaert:2005jf,BCL} may be
summarized in the following theorem in a form which suggests
itself a conjecture for an arbitrary Young diagram.

\vspace{1mm}\noindent{\bfseries{Deformations of the algebra:}}
\textit{For a collection of gauge fields $\varphi^a_Y$
($a=1,\ldots,N$) labeled by a f\underline{ixed} Young diagram $Y$
with three columns or less, the non-Abelian interaction problem
does not possess any non-trivial solution if the Young diagram $Y$
is made of more than one row. In the completely symmetric tensor
case, at first order in some smooth deformation parameter,
Poincar\'e-invariant deformations of the (Abelian) gauge algebra exist.
The deformed gauge algebra may always be assumed to be closed off-shell.
Two cases arise depending on the parity-symmetry property of the first-order deformation.
\begin{itemize}
    \item[(i)]
  The first-order {\normalfont{parity-invariant}}
  deformations of the gauge algebra are in one-to-one
  correspondence with the structure constant tensors
  $C^a{}_{bc}=(-)^s C^a{}_{cb}$ of an (anti)commutative internal algebra,
  that may be taken as deformation parameters;
  \item[(ii)]
  The first-order {\normalfont{parity-breaking}} deformations of the gauge algebra are
  characterized by structure constant tensors
  $C^a{}_{bc}=(-)^s (\delta^n_3-\delta^n_5)C^a{}_{cb}$,
  where $n$ is the space-time dimension and $s>1$.
\end{itemize}
}\vspace{1mm}

In other words, one may conjecture that there exists no solution
to the non-Abelian interaction problem for any finite collection
of mixed symmetry gauge fields (at least for fixed symmetry
properties). Therefore, from now on we focus on the case of a
collection $\varphi^a_{\mu_1\ldots\mu_s}$ ($a=1,\ldots,N$) of
completely symmetric tensor gauge fields with fixed spin $s$. We
also review the lower spin cases $s=1,2$ in order to compare them
with the higher-spin case $s=3$ and look for similarities or
novelties.

\vspace{1mm}\noindent{\bfseries{Deformations of the
transformations:}} \textit{For a collection
$\varphi^a_{\mu_1\ldots\mu_s}$ ($a=1,\ldots,N$) of completely
symmetric tensor gauge fields with fixed spin $s=1,2,3,$ the most
general Poincar\'e and  p\underline{ari}ty\underline{-invariant}
gauge transformations deforming the gauge algebra at first order
in the structure constants are equal to, up to gauge
transformations that either are trivial or do not deform the gauge
algebra at first order,
\begin{description}
  \item[$s=1$] the Yang-Mills gauge transformation $\d \varphi^a_\mu=\partial_\m \varepsilon^a-\,C^a{}_{bc}\, \varphi^b_\m\,
  \varepsilon^c$;
  \item[$s=2$] the ``multi-diffeomorphisms"
  $\d\varphi^a_{\m\n}=2\,\partial^{}_{(\mu}\varepsilon^a_{\nu)}\,-
  \,C^a{}_{bc}\,\eta^{\r\s}\,(
  \,2\partial^{}_{(\m}\varphi^b_{\n)\r}-\partial^{}_\r\varphi_{\m\n}^b\,)\,\varepsilon^c_\s
  + \co(C^2)$;
  \item[$s=3$] the spin-3 gauge transformations decomposing into two
categories. More precisely, the structure constant tensor $C$
splits into $f$ and $g$ and the gauge transformations are of the
schematic form
$$\delta \varphi^a_{\m\n\r} = 3\,\pa^{}_{(\m}\varepsilon^a_{\n\r)}+f^a{}_{bc}\,(\partial\varphi^b
\partial\varepsilon^c)_{\m\n\r}+\,g^a{}_{bc}\,(\partial^2\varphi^b
\partial^2\varepsilon^c )_{\m\n\r}\,+\co(C^2)\,,$$
where the structure constant tensor $g^a{}_{bc}$ vanishes in
space-time dimension $n<5$.
\end{description}
}\vspace{1mm}

In length units, the coupling constants $C^a{}_{bc}$ have
dimension $n/2+s-3$, except for $g^a{}_{bc}$ which has dimension
$n/2+2$. The spin-three deformation associated with the tensor
$f^a{}_{bc}$ was obtained in \cite{Berends:1984wp} while the
spin-three deformation corresponding to $g^a{}_{bc}$ was obtained
in \cite{Bekaert:2005jf}. \vspace{.2cm}

\textit{In the spin-2 and spin-3
p\underline{ari}ty\underline{-breakin}g cases, the first-order
deformations of the gauge transformations schematically read (for
the detailed expressions, see the second ref. of
\cite{Boulanger:2000rq} and ref. \cite{BCL})
\begin{description}
\item[$s=2$] \quad $\d \varphi^a_{\m\n}=2\,\partial^{}_{(\mu}\varepsilon^a_{\nu)} +
\delta^n_3\,\widetilde{f}^{a}_{~bc}\, \varepsilon^{\m\n\r}(\pa
\varphi^b\pa \varepsilon^c)_{\m\n\r} +
\delta^n_5\,\widetilde{g}^{a}_{~bc}\, \varepsilon^{\m\n\r\s\l}
(\pa \varphi^b\pa \varepsilon^c)_{\m\n\r\s\l}\,$,
\item[$s=3$] \quad $\d \varphi^a_{\m\n\r}=3\,\partial^{}_{(\mu}\varepsilon^a_{\nu\r)} +
\delta^n_3\,\widehat{f}^{a}_{~bc}\, \varepsilon^{\m\n\r} (\pa
\varphi^b \varepsilon^c+\varphi^b \pa\varepsilon^c)_{\m\n\r} +
\delta^n_5\,\widehat{g}^{a}_{~bc} \,\varepsilon^{\m\n\r\s\l}
(\pa^2 \varphi^b \pa \varepsilon^c)_{\m\n\r\s\l}$\,.
\end{description}}

\section{Non-Abelian cubic vertices}

In order to provide an intrinsic characterization of the
conditions on the constant tensors characterizing the
deformations, let us start by briefly reviewing some basics of
abstract algebra.

\subsection{Algebraic preliminaries}

Let $\cal A$ be a real algebra of dimension $N$ with a basis
$\{T_a\}$. Its multiplication law $\ast:{\cal A}^2\rightarrow \cal
A$ obeys $a\ast b=(-)\,b\ast a$
if it is {\it (anti)commutative}, which is equivalent to the fact
that the structure constant tensor defined by $T_b\ast
T_c=C^a{}_{bc}\,T_a$ is (anti)symmetric in the covariant indices:
$C^a{}_{bc}=(-)\,C^a{}_{cb}$. Moreover, let us assume that the
algebra $\cal A$ is an Euclidean space, {\it i.e.} it is endowed
with a scalar product $\langle\,\,,\,\rangle:{\cal A}^2\rightarrow
\mathbb R$ with respect to which the basis $\{T_a\}$ is
orthonormal, $\langle\,T_a\,,\,T_b\,\rangle=\d_{ab}$. For an
(anti)commutative algebra, the scalar product is said to be {\it
invariant} (under the left or right multiplication) if and only if
$\langle\,a\ast b\,,\,c\,\rangle=\langle\,a\,,\, b\ast c\,\rangle$
for any $a,b,c\in \cal A\,$, and the latter property is equivalent
to the complete (anti)symmetry of the trilinear form
$$C:{\cal A}^3\rightarrow{\mathbb R}:(a,b,c)\mapsto
C(a,b,c)=\langle\,a\,,\,b\ast c\,\rangle$$ or, in components, to
the complete (anti)symmetry property of the covariant tensor
$C_{abc}:=\d_{ad}\,C^d{}_{bc}$. For that reason, the former
algebras are said to be {\it (anti)symmetric}.
An anticommutative algebra satisfying the Jacobi identity
$a\ast(b\ast c)+b\ast(c\ast a)+c\ast(a\ast b)=0$ is called a Lie
algebra and the product $\ast$ is called a Lie bracket. In
components it reads $C^e{}_{d[a}C^d{}_{bc]}=0$. Furthermore, the
Killing form of a (compact) semisimple Lie algebra endows it with
an (Euclidean) antisymmetric algebra structure. Eventually, an
algebra is said to be associative if $a\ast(b\ast c)=(a\ast b)\ast
c$ which, for (anti)commutative algebras, reads in components
$C^d{}_{b[c}C^e{}_{a]d}=0\,$. For anticommutative algebras, the
associativity is much stronger than the Jacobi identity.

\subsection{Cubic vertices}

An important physical question is whether or not these first-order
gauge symmetry deformations possess some Lagrangian counterpart.
The following theorem provides a sufficient condition
\cite{Barnich:1993pa,Boulanger:2000rq,Bekaert:2005jf,BCL}.

\vspace{1mm}\noindent{\bfseries{Cubic vertices:}} \textit{Let  the
constant tensor $C_{abc}:=\d_{ad}C^d{}_{bc}$ be completely
(anti)symmetric, then the non-Abelian interaction problem for a
quadratic local action $S_0[\varphi^a_{\mu_1\ldots\mu_s}]$ (rank
$s$ fixed) admits first-order solutions which are local
functionals $C_{abc}\,S^{abc}_1[\varphi^a_{\mu_1\ldots\mu_s}]$
such that the deformation
$S=S_0\,+\,C_{abc}\,S^{abc}_1\,+\,\co(C^2)$ is invariant under the
aforementioned gauge transformations at first order in $C$. They
\begin{description}
  \item[s=1] are equal to the Yang-Mills cubic vertex $S_1[\varphi^d_{\mu}]=C_{[abc]}\int d^nx\,\,\partial^{[\m}\varphi^{\n]\,a}\varphi_\m^b
  \varphi_\n^c\,$.
  \item[s=2] decompose as a sum
  $S_1[\varphi^d_{\mu\n}]=C_{(abc)}R^{abc}\,+\,\d^n_3\,\widetilde{f}_{(abc)}\widetilde{U}^{abc}\,+\,\d^n_5\,\widetilde{g}_{[abc]}\widetilde{V}^{abc}$
of cubic functionals respectively containing two, three and three
derivatives.
  \item[s=3] decompose as a sum
  $S_1[\varphi^d_{\mu\n\r}]=f_{[abc]}\,S^{abc}\,+\,g_{[abc]}\,T^{abc}\,+
  \,\d^n_3\,\widehat{f}_{[abc]}\widehat{U}^{abc}\,+\,\d^n_5\,\widehat{g}_{(abc)}\widehat{V}^{abc}$ of cubic functionals respectively containing three, five, two and four
  derivatives.
\end{description}
The vertices in the first-order deformations are determined
uniquely by the structure constants, modulo vertices that do not
deform the gauge algebra. Moreover, the (anti)symmetry of the
internal algebra is not only a \underline{su}ff\underline{icient}
but also a \underline{necessar}y requirement for all known cases
(i.e. this issue is open only for the spin-three case with the
deformation associated with $g^a{}_{bc}$).}\vspace{1mm}

The first-order covariant cubic deformation
$S^{abc}[\varphi^d_{\m\n\r}]$ is the BBvD vertex
\cite{Berends:1984wp}. We do not know yet whether the antisymmetry
condition on the structure constant $g^a{}_{bc}$ is actually
necessary or not for the existence of a consistent vertex at first
order but, looking at all other cases, it seems very plausible.
One may thus conjecture that, for any fixed helicity $s$, the
existence of a local Lagrangian counterpart to the non-Abelian
gauge symmetries requires that the structure constants define an
(anti)symmetric internal algebra. The fact that the internal
algebra is Euclidean is hidden in the fact that in the free limit
the action is a {\it positive} sum of quadratic {\it ghost-free}
actions.

\section{Consistency at second order}

Of course, the next issue is whether the first order deformations
can be pushed further or if they are obstructed. Consistency of
the gauge algebras (only by itself and already at second order)
constrains rather strongly the parity-invariant possibilities
\cite{Barnich:1993pa,Boulanger:2000rq,Bengtsson:1983bp,Noether,Bekaert:2005jf}.
In the spin-2 and spin-3 parity-breaking cases, the issue is more
subtle and a detailed comparative discussion is given in the
conclusion of \cite{BCL}.

\vspace{1mm}\noindent{\bfseries{Consistency of the algebra:}}
\textit{At second order in $C^a{}_{bc}$, the parity-invariant deformation
of the gauge algebra can be assumed to close off-shell without loss of
generality, and for $s=1,2,3$ it is not obstructed if and only if
the structure constants $C^a{}_{bc}$ define an internal algebra
which is
\begin{description}
  \item[s=1] a Lie algebra.
  \item[s=2] an associative algebra.
  \item[s=3] an associative algebra for $f^a{}_{bc}$, or if the space-time
  dimension is equal to $n=3$.
\end{description}
}

We emphasize that the existence of a cubic vertex corresponding to
the non-Abelian gauge transformations was not necessary to derive
this theorem. In order to combine the latter result with the
former one for the existence of a Lagrangian counterpart, one may
use the following well-known lemma.\footnote{The proofs are
elementary and were given in the corresponding references.}

\vspace{1mm}\noindent{\bfseries{Lemma:}} \textit{If a
finite-dimensional (anti)commutative (anti)symmetric Euclidian
algebra is associative, then it is the direct sum of
one-dimensional ideals.}\vspace{1mm}

This lemma leads to stringent restrictions on the deformations
which are consistent till second order.

\vspace{1mm}\noindent{\bfseries{Corollary:}} \textit{For a
collection $\varphi^a_{\mu_1\ldots\mu_s}$ ($a=1,\ldots,N$) of
completely symmetric tensor gauge fields with fixed rank
$s=1,2,3,$ the non-Abelian parity-invariant interaction problem is
such that the deformed gauge algebra is
\begin{description}
  \item[s=1] a finite-dimensional internal Lie algebra endowed with an antisymmetric algebra
  structure. For semi-simple compact Lie
  algebras, the scalar product is naturally identified with the Killing
  form.
  \item[s=2] given by the direct sum of diffeomorphism (i.e. vector field)
 Lie algebras.
  \item[s=3] inconsistent if $f^a_{bc}\neq 0$ and $n>3$.
\end{description}
}\vspace{1mm}

To conclude, the recent (modest but exciting) observations on the
spin-three non-Abelian interaction problem are that, at second
order at the level of the gauge {\it algebra}, the new
deformations corresponding to the structure constants $g_{[abc]}$
\cite{Bekaert:2005jf} and $\widehat{g}_{(abc)}$ \cite{BCL} both
pass the consistency requirement where the BBvD vertex fails, and
that the BBvD gauge symmetries are not obstructed in
three-dimensional flat space-time (this new result is proper to
the present paper). Unfortunately, we do not know yet whether
there exist second-order gauge {\em{transformations}} that are
consistent at this order.

\section*{Acknowledgments}
X.B. is grateful to the organizers of the network meeting for
giving him the opportunity to present this work in Corfu. He also
thanks the Albert Einstein Institut, the Werner Heisenberg
Institut and especially the Institut des Hautes \'Etudes
Scientifiques for hospitality.

\noindent The work of S.C. is supported in part by the
``Interuniversity Attraction Poles Programme -- Belgian Science
Policy'' and  by IISN-Belgium (convention 4.4505.86). Moreover,
X.B. and S.C. are supported in part by the European Commission FP6
programme MRTN-CT-2004-005104, in which S.C. is associated to the
V.U.Brussel (Belgium).

\end{document}